**On the nonexistence of a liquid-gas critical point and the existence of a supercritical mesophase**

Leslie V. Woodcock

Manchester Interdisciplinary Biocentre, University of Manchester,
Manchester M1 7DN, United Kingdom

**ABSTRACT:**       We extend previous investigations into the thermodynamics of liquid state boundaries by focusing on the origins of liquid-gas criticality. The singular point hypothesis of van der Waals is re-examined in the light of recent knowledge of the hard-sphere percolation transitions and further analysis of simulation results for the supercritical properties of the square-well fluids. We find a thermodynamic description of gas-liquid criticality that is quite different from both van der Waals hypothesis and modern mean-field theory. At the critical temperature ($T_c$) and critical pressure ($p_c$), in the density surface $\rho(p,T)$, there is no critical point. Using tabulations of experimental $\rho(p,T)$ data,  for supercritical argon, and also water, as examples, at $T_c$   a liquid phase coexists with a vapor phase determined by percolation transition densities.   In the $\rho(p,T)$ surface, there is a line of critical coexistence states of constant chemical potential at the intersection of two percolation loci in the p-T plane. For temperatures above this line, there exists a supercritical mesophase bounded by percolation transition loci. Below the line of critical states there is the familiar subcritical liquid-vapor two-phase coexistence region. Unlike the hypothetical van der Waals critical point, the line of critical states complies with Gibbs phase rule.

## INTRODUCTION

In a previous article on this subject[1] it was found that two percolation transitions of available volume (PA), and extended volume (PE), play a central role in determining liquid state boundaries. The concept of a liquid-gas critical point has a long history in the physical chemistry of fluid phase thermodynamic equilibria and indeed, more generally across condensed matter physics. A critical temperature (T), above which a gas cannot be liquefied by application of pressure (p), was discovered for carbon dioxide by Andrews in 1861.[2] The first to suggest the existence of a critical point. was van der Waals, in 1873.[2,3] His hypothesis of a singularity on the density surface $\rho(p,T)$, as represented by the node of a cubic equation, has since remained the unreservedly accepted thermodynamic description of liquid-gas critical phenomena.[4,5]

The failure of the Mayer virial expansion to converge onto the thermodynamic pressure as a function of density $p(\rho)$ at the percolation transition density $\rho_{pa}$ has implications for both the low temperature liquid state limit, on supercooling, and also for the high temperature limit of equilibrium liquid-vapor coexistence. Percolation transitions were seen to be the originating thermodynamic states of the coexistence properties of the square-well fluid. Here we confirm the role of hard-sphere percolation transitions in square-well criticality, and then explain how the previous result, that the critical "point" is actually a line in the $\rho(p,T)$ density surface, extends to liquid argon and water, and hence also all other real liquids.

Despite an extensive literature on theory of critical phenomena in lower dimensions, and also lattice gases for which critical-point properties can be evaluated analytically[4], in the 140 years since van der Waals, there have been no theoretical developments to justify the singular point description of liquid-gas criticality.[2-5] We reveal more evidence, from both new and old simulation results for square well fluids, together with further analysis of real liquids, argon and water, all of which show that the "critical point", as envisaged by van der Waals, does not exist as such.

When van der Waals wrote his renowned thesis on the theory of the critical point in 1873, he would not have been aware of Gibbs work on thermodynamic equilibria,[6] also published in 1873. The van der Waals critical point does not comply with Gibbs phase rule. Its existence is based upon a hypothesis rather than a thermodynamic definition. Moreover, no one has ever succeeded in measuring a critical density of an atomic or molecular fluid directly. Liquid-gas critical densities are only obtained experimentally indirectly by an extrapolation of a mean of the two coexisting densities of liquid and vapor using the law of rectilinear diameters. The existence of a critical point singularity in the $\rho(p,T)$ density surface, however, has not been questioned until very recently.[1]

Gibbs defined surfaces of thermodynamic state functions, and subsequently explained all 1st -order thermodynamic phase transitions. A point on the $\rho(p,T)$ surface can only be defined thermodynamically where two lines intersect. For example, in a single phase region, a state point with two degrees-of-freedom (F) is the intersection of an isobar and an isotherm. A state point in a two-phase region (F=1), is the intersection of either an isotherm or an isobar with a coexistence line. A coexistence line is defined by the intersection of Gibbs chemical potential loci for respective phases. There are no conceptual problems with the definition of the triple-point (F=0); it is the intersection of the liquid-vapor coexistence line and the solid-vapor coexistence line in the p-T plane. By contrast, there is no thermodynamic definition of the "critical point" of van der Waals; indeed if at $T_c$ $(dp/d\rho)_T = 0$, the state has one phase and one degree of freedom in an apparent contradiction of Gibbs phase rule.

In the following sections, we present evidence for the non-existence of a van der Waals singularity, but instead, the existence of a line of critical states, which resolves this Gibbs noncompliance conundrum. Analysis of model square-well fluids, phases of argon, and even also water and steam, all provide the compelling evidence that at a critical $T_c$ and $p_c$ there exists a 2-phase coexistence line of critical states connecting the densities of two percolation transitions defined by the intersection of bonded-cluster percolation (PB) and available-volume percolation (PA) loci in the p-T plane.

## MEAN-FIELD THEORY

The main motivation of van der Waals was to understand and describe the liquefaction experiments of Andrews on carbon dioxide[2,3]. He assumed that the equation for the isothermal change of pressure as a function of volume must contain an inflection at a "critical point". This was necessary in order to give two roots for volume at temperatures below the critical isotherm, and a single root for volume at temperatures above the critical isotherm. Thus, he proposed that the equation-of-state for the pressure of a fluid should be cubic in terms of volume. This is essentially a mathematical parameterization of a physical hypothesis that the liquid–vapor coexistence line, in the $\rho(p,T)$ surface diminishes with increasing temperature all the way to a point, and disappears.

Accordingly, van der Waals obtained his renowned equation of state for gases and liquids by simply taking the ideal gas equation-of-state and, treating molecules as impenetrable spheres (diameter $\sigma$). He represented the pressure of a hard-sphere fluid as a function of absolute temperature (T) and volume (V) by simply replacing the volume by (V-b). The constant b is the excluded-volume on collision of two spheres $(2\pi\sigma^3/3)$. He then deducted from the pressure a term for the mean-field attractive energy proportional to density squared.

$$p = Nk_BT/(V-Nb) - a\,\rho^2 \qquad (1)$$

where the density $\rho = N/V$.

Van der Waals equation (VDW) is useful at low densities, but we can see from the isotherms, plotted here in **Figure 1a,** in the high-temperature limit, his equation-of-state for hard-spheres is a very poor representation. At temperatures below the critical point, moreover, equation (1) becomes unphysical. For all $T < T_c$ it predicts negative compressibility, and at temperatures below $0.85T_c$ negative pressures are obtained. Nevertheless, when van der Waals equation for the hard-sphere fluid pressure, p= $Nk_BT/(V-Nb)$, is expanded in powers of density, we see that the first two terms are exact, b is the second virial coefficient. The mean-field $a\rho^2$ term, moreover, can also be shown to be correct for certain types of model attractive molecules in the low density limit.[4,5]   Equation (1), is essentially an accurate theory of non-ideal gases; but it tells us little, if anything, about liquids or liquid-vapor coexistence.

When an accurate equation for the hard-sphere fluid pressure  is substituted into van der Waals equation (1), a much improved equation is obtained, which is referred to as the "augmented van der Waals equation" (AVDW)

$$p = p_{hs} (\rho,T) - a \, \rho^2 \qquad\qquad (2)$$

where $p_{hs}$  is the hard-sphere fluid pressure. Van der Waals attractive constant (a) is deemed to be independent of T.  This implies that the structure of the hard-sphere fluid, and a van der Waals model fluid, are everywhere the same.  Some isotherms for the AVDW equation are plotted in **Figure 1b**. There is a big improvement on VDW; the high temperature limit is now correct. A critical point singularity remains; the density is shifted from 0.16 to 0.25, i.e. closer to typical literature values of molecular-reduced critical densities of real simple liquids.

Equations (1) and (2) can also be derived for a model pair potential to gain a better understanding of the physical approximations. In perturbation theory the change in free energy of the fluid is expanded in powers of density. The zeroth term is the reference hard-sphere fluid; in AVDW only the first-order term is retained. Thus we have

$$A = A_0 + ½ \, N \, \Phi \qquad\qquad (3)$$

where A is the excess Helmoltz free energy; $A_0$ is the free-energy of the hard-sphere reference fluid and $\Phi$ is the "mean-field" potential energy of attraction per particle. Assuming pairwise additivity, the attractive potential can be expressed as an integral over the pair distribution function

$$\Phi = \int (N/V) \, 4\pi r^2 \phi(r) \, g_0(r) \, dr \qquad\qquad (4)$$

Where $g_0(r)$ is pair distribution function of the reference hard-sphere fluid and $\phi(r)$ is the intermolecular pair potential.  In van der Waals approximation, there is total randomness for $r > \sigma$ then, $g_0(r) = 1$ everywhere, whereupon the attraction constant is

$$a = \int 2\pi \, \phi(r) \, r^2 dr \qquad\qquad (5)$$

The simplest pair potential that represents van der Waals theory of the hard-sphere repulsion, and the mean-field attraction given by equation (5) is the square-well. Thus, provided $\phi(r)$ goes to zero at some distance, in the first-order of perturbation, a is a molecular constant independent of temperature (T); below we test this approximation for a square-well fluid with a wide well, i.e. a favorable case for approximations in equations (3), (4) and (5).

## SQUARE-WELL FLUIDS

The square-well Hamiltonian is the simplest model of a van der Waals molecular fluid and has been widely investigated as a test of the AVDW equation. The pair potential for separation r is

$$\phi(r) \qquad = 0 \text{ for} \qquad r > \lambda \, \sigma$$

$$= -\varepsilon \quad \text{for} \qquad \sigma > r > \lambda \, \sigma$$

$$= \infty \text{ for} \qquad r < \sigma$$

where $\sigma$ is the hard-sphere diameter, $\varepsilon$ is the square-well depth, and $\lambda\sigma$ is the range.

Thermodynamic properties of square-well fluids, i.e. for various values of $\lambda$, can be calculated from molecular dynamics (MD) or Monte Carlo (MC) computer simulations. There are a number of reported MC and MD investigations of the coexistence properties of square-well fluids [7-11], ranging from studies of percolation thresholds,[7] theoretical studies of mean-field approximations,[8] MC Gibbs ensemble calculations[9] and MD investigations of liquid-vapor coexistence.[10,11] All of the computer simulation studies are in the well-width range $\lambda < 3$. It is believed by some [5,8-11] that when the square-well width becomes large, equation (2) will become exact.

High-performance computing has recently been applied to the very accurate determination of the virial equation-of-state for hard spheres and also the thermodynamic equation of state.[12] We now know that the hard-sphere fluid equation of state is not continuous all the way to freezing; there exists two percolation transitions.[13] The excluded volume percolation transition occurs at the density $\rho_{pe}$=0.0785, above which clusters of spheres of the excluded volume first span the system. The available volume percolation transition occurs at the density $\rho_{pa}$=0. 537; above which the volume available to an additional sphere ceases to percolate the system.

New MD results (**Figure 1c**) for the square-well fluid $\lambda$=5, show that this is not so. A summary of values obtained to date for the mean pressures of the square-well fluid ($\lambda$=5) is given in **Table 1**; the original isotherms are plotted in **Figure 2**. The two percolation transition densities are indicated by dashed lines. The critical isotherm is estimated to lie between the two isotherms at $T_c^*$ = 42; using this value, the critical coexistence pressure is obtained by interpolation between the two isotherms $T^*$=40 and $T^*$=45 and averaging the pressure between two percolation transition densities. The estimate critical pressure is $P_c^*$= 0.0579 based only upon the MC pressures for the finite system N= 6912. We do not have a reliable estimate of the N-dependence at present. The infinite-system pressure is likely to be slightly higher, i.e. closer to the equation (2) value of $P^*_c$ (AVDW) = 0.0873.

At percolation transitions, the state dependence of density fluctuations may change, in which case the compressibility $(dV/dp)_T$ will exhibit a second or higher-order phase transition. There is also a change in the form of the van der Waals perturbation at these transitions which is discernable from the isotherms for example, of the square-well fluid $\lambda=5$ as shown in **Figure 2**. The strength of the discontinuity is seen to increase with decreasing temperature. Mean-field approximations such as the AVDW equation, which may only be expected to be representative up to the first percolation transition for square-well fluids of long range, miss the essential physics of the percolation transitions, and hence also the critical condensation behavior.

A comparison of the AVDW equation (2) (**Figure 1b**) with similar MD isotherms for the square-well fluid ($\lambda=5$) in **Figure 1c** confirm the deficiencies in the mean-field approximation equation (2). The first observation is a change in slope of $p(\rho)$ at the density of the hard-sphere available volume percolation transition $\rho_{pa}$. The second observation is that at a critical temperature there is an apparent coexistence line of constant pressure, and hence also chemical potential since $dp = 0$ as $\rho$ increases, from a dilute gas state to the percolation transition density $\rho_{pa}$. There is no evidence of the van der Waals critical point seen in **Figures 1a and 1b**.

Insight into the reason for the failures of the AVDW equation can be gleaned from the analysis of the variation in $p_{hs}$ - $p_{sw}$ computed from MD calculation for the fluid $\lambda=5$ (**Figure 3**). The isotherms have been fitted with a quadratic trendline to test the obedience to the AVDW first order perturbation equation (2). The $\rho^2$ dominance is a good approximation; the very small linear correction is due to the fact that in the linear term, both hard-sphere pressure ($p_{hs}$) and square-well pressures ($p_{sw}$) obey the ideal gas law, they have the same dependence of $p$ as $\rho \rightarrow 0$. More importantly, the mean-field constant (a) has a dependence upon T, unlike the constant in equation (2), that strengthens as $T_c$ is approached. It is this dependence that signals the complete failure of the approximation to reflect the intersection of PE and PA at $T_c$. Far from the theory predicting the physical properties of any critical phenomena, the "mean-field" critical-point of equation (2) is basically a mathematical "catastrophe", devoid of physical reality.

Notice in **Figure 1c** that the density of the vapor coexistence is in the region of the excluded volume hard-sphere percolation transition ($\rho_{pe}$ =0.0785) and also in **Figure 2 and 3** that the slopes of all the $\Delta p(\rho)$ isotherms increase rather sharply around $\rho_{pe}$. It may be inferred that the coexisting densities of square-well fluids, at least for $\lambda > 2$, and as $\beta_c =1/T^*_c \rightarrow 0$, that the vapor and liquid coexisting densities are $\rho_{pe}$ and $\rho_{pa}$ of the hard sphere fluid.

If the mean-field approximation is inapplicable to long-range square-well fluids, what happens to the coexistence region when the square-well becomes narrow, as in real molecules perhaps? There is now another percolation transition to consider, which occurs at an intermediate density for $\lambda < 2$, when a connected cluster of "bonded" molecules within the square-well attractive range, first spans the whole system.[7] $\rho_{pb}(\lambda)$. For the high-temperature limit of all square-well fluids, i.e. the hard-sphere fluid, the available volume percolation transition is a constant independent of $\lambda$ . For the special square-well case of $\lambda=2$, then $\rho_{pe}$(HS) and $\rho_{pb}(\lambda=2$: SW) T$\rightarrow\infty$ are the same.

This dependence $\rho_{pb}(\lambda)$ has been computed by Heyes et al. in the high temperature limit, i.e. for the hard-sphere reference model of square-well fluids.[7] We also know the dependence of the critical temperature on λ for square-well fluids; an equation for $T^*_c(\lambda)$, where $T^* = k_B T/\varepsilon$ and $k_B$ is Boltzmann's constant, is given by Scholl-Passenger et al.[8] Combining the two functions of λ, we obtain a simple expression for the dependence of the extended-volume percolation transition density upon the critical temperature $T_c^*$ of square-well fluids.

$$\rho_{pe}(T_c^*) = 0.125\ T_c^{*\ -3/4} \qquad (6)$$

All three percolation loci are shown in **Figure 4** alongside the coexistence curves obtained from Vega et al.[9] and also Elliott and Hu,[10] the direct coexistence results of Benavides et al.[11] (λ = 3) , and present Monte Carlo simulation results for λ = 1.005 and λ=5. Here, no assumptions are made regarding the form of the "inaccessible region" near a "critical point", or its exponents. The curves are actually well-represented by quadratics. **Figure 4** also shows the lowest density point of the liquid and the highest density point of the vapor phase. The percolation transitions follow the lowest and highest densities of the liquid and vapor in coexistence, respectively, obtainable in Gibbs ensemble Monte Carlo simulations, and also from direct MD simulations[11], also shown in **Figure 4**.

The maximum coexisting vapor density follows the loci that one expects, not for the hard-sphere excluded-volume percolation (λ=2), but the extended-volume, or equivalently, the bonded cluster percolation transition of the square well fluid. The critical densities, within the rather wide uncertainty that they have been obtained, are intermediate of the two percolation transitions. At $T_c$, the vapor at, or close to the density $\rho_{pe}$, and the liquid at $\rho_{pa}$ must have the same chemical potential as the gradient in chemical potential ($d\mu = Vdp = dp/\rho$) approaches zero between $\rho_{pe}$ and $\rho_{pa}$ as the density fluctuations diverge. For a large well-width, e.g. λ=5, the coexisting vapor density is close to the hard-sphere percolation density[7,13] $\rho_{pe} = 0.0785$. At very narrow well-width, almost at the sticky sphere limit (λ=1), $\rho_{pe}$ (λ=1.005) and $\rho_{pa}$ (λ=1.005) are quite close and the critical tie-line is relatively narrow.

An interesting question is why is the vapor density apparently determined by the constant hard-sphere $\rho_{pe}$ (=0.0785) for square-well fluids λ >2, whereas it is determined by the value of $\rho_{pe}(\lambda)$ for λ < 2? We conjecture that the answer is as follows. Square-well fluids have only one available-volume percolation transition, but they have two extended volume percolation transitions, one defined by the hard-sphere excluded volume at the distance 2σ, and a second defined by the square well width λ. It is the percolation transition of higher density, and hence also the higher pressure, that will first intersect with the available volume percolation pressure $p_{pa}(T)$, and thereupon cause the 1st-order phase transition.

## FLUID PHASES OF ARGON

Next, we consider the simplest real classical fluid, argon. The experimental literature of p-V-T data, up to and well beyond the critical temperature, has been extensively measured and tabulated to a precision of 6-figure accuracy.[14] The numerical data for the critical isotherm and seven supercritical isotherms are taken directly from the Gilgen tables, and reproduced in **Figure 5** as plotted from an EXCEL spreadsheet.

The second-order percolation discontinuities in dp/dρ for supercritical isotherms of argon can be clearly identified. The pressure varies linearly with density in the region between the percolation transitions which are represented by near-vertical loci lines in Figure 5. The percolation transition pressures become the same at the mass densities around $\rho_m$ = 600 and 450 kg/ m$^3$, for liquid and vapor respectively, at the critical pressure 4.8MPa.

At the critical temperature, there is a 2-phase coexistence line between the densities of two percolation transitions. By analogy with the triple point, $T_c, p_c$ is found to be a "double point" with a single degree of freedom in the p-T plane where the percolation transition pressure loci intersect, as shown in **Figure 6** for fluid argon. At $T_c$, each state-point corresponds to a different density, and since $(Vdp)_T = d\mu = 0$, there is a connecting line of states at $T_c, p_c$ of constant Gibbs chemical potential ($\mu$). The liquid-vapor coexistence line is drawn from the experimental p-V-T data of Gilgen et al.[15] The data points for the two percolation transitions are obtained from the discontinuity in the slope of the p($\rho$) isotherms, for seven supercritical isotherms. Using this information, and knowledge of the percolation transitions for the high-temperature limit of argon, if it is represented by a Lennard-Jones model, we can obtain a description of the entire fluid phase diagram.

The available volume ($V_a$) percolation transition (PA) occurs at the density ($\rho_{pa}$) at which the volume accessible to any single mobile atom, in static equilibrium configuration of all the other atoms, percolates the whole system. It is related to the Gibbs chemical potential ($\mu$) by the equation

$$\mu = -k_B T \log_e V_a \qquad\qquad (7)$$

The bonded-cluster percolation transition (PB), is the same percolation transition, previously referred to for square-well fluids[1] as the extended-volume percolation transition. For real molecules we must now distinguish between the excluded volume percolation transition (PE), which is also a molecular cluster system-spanning transition, and the bonded-cluster percolation transition (PB).

At percolation transitions, thermodynamic state functions can change form due to sudden changes in state-dependence of density and/or energy fluctuations. For the hard-sphere fluid, PA is a very weak, but definite, higher-order thermodynamic phase transition.[13] Purely repulsive potential models have a gas-like region and a liquid-like region on either side of PA.

We do not presently know whether PE has a thermodynamic status for the hard-sphere fluid. When an attractive perturbation is added, however, both percolation transitions gain strength as temperature is reduced. This gives rise to 2$^{nd}$-order thermodynamic phase transitions, in which there are discontinuities in second derivatives of chemical potential with respect to temperature or pressure, notably: isothermal compressibility $(d_2\mu/dp^2)_T$, heat capacity $(d_2\mu/dT^2)_p$ and thermal expansivity $(d_2\mu/dpdT)$ all of which undergo some degree of change across percolation transitions.

The bonded-cluster percolation transition (PB) occurs when atoms bonded together, within a given characteristic distance, around the minimum in the pair potential, first begins to span the system. Unlike PA, PB manifests itself more in the temperature derivatives of the chemical potential which are determined by fluctuations in the energy, rather than density. Thus we are more likely to see lines of discontinuity, showing apparent maxima or minima, in the 2nd- order properties heat capacity and thermal expansivity.

The p-T plane projection in **Figures 6** shows only the two percolation transitions that determine the critical coexistence. Atomic fluids have only one available-volume percolation transition (PA), but of the other two percolation transitions, the bonded-cluster percolation transition (PB) occurs at a much higher density than the excluded volume percolation transition (PE). Thus the bonded-cluster percolation intersects the available volume percolation line first at the higher temperature and pressure to effect the first-order phase transition. The percolation transition points are obtained from the isotherms by noting that in the supercritical mesophase $p(\rho)$ isotherms are linear, whereas for both gas-like region and liquid-like region, the $p(\rho)$ has power-law forms. The point of deviation is a 2nd-order phase change, in energy and density fluctuations, and which defines the percolation transition pressure.

It is interesting that the percolation transition pressures for argon are linear functions of temperature. This is consistent with a simple scaling law. In a perturbation model of "hard-sphere argon", $p_{pe}\sigma^3/k_B T_{pe}$ and $p_{pa}\sigma^3/k_B T_{pa}$ are the reference hard-sphere constant values, so both $p_{pe}$ and $p_{pa}$ will vary linearly with T, with $p_{pa}$ having the higher slope. At present, we do not have information on the excluded volume percolation transition in the Lennard-Jones (L-J) fluid, and its effect, if any, at lower temperatures, i.e. on intersecting the vapor phase coexistence line at a temperature between $T_c$ and the triple point.

To obtain a general simple-fluid phase diagram, we use the L-J potential with the scalable energy ($\epsilon$) and distance corresponding to the diameter of a hard sphere reference fluid where $r_0$ is in dimension of the distance of zero force at $\epsilon = -1$.

$$\phi(r) = \epsilon \left[ (r/r_0)^{-12} - 2(r/r_0)^{-6} \right] \qquad (8)$$

The phase diagram is presented in **Figure 7,** and has been constructed follows. Beginning with the raw data from the Gilgen tables,[14,15] the experimental mass densities at $T_c$ can be converted to reduced number densities for comparison with the known percolation transitions of the hard-sphere reference fluid. Using a Lennard–Jones pair potential for argon[5] ; $\epsilon/k_B = 120K$ and $\sigma = 3.405 \times 10^{-10}m$ .The reference hard-sphere diameter corresponding to zero-force is $r_0 = 2^{(1/6)} \sigma = 3.822 \times 10^{10}m$; taking Avagadro's number N= $6.0228 \times 10^{23}$ and argon atomic mass = 39.948, the data points of Gilgen have been converted to L-J units, with reduced number density $\rho = Nr_0^3/V$.

Nobody has ever measured a critical density directly; this is well illustrated from the experimental measurements of the argon liquid vapor coexistence densities by Gilgen et al.[15] The highest temperature for which they report both coexisting vapor and liquid densities is 150.61. They use the law of rectilinear diameters to obtain a "critical point" temperature 150.69, and a critical density 535.6 kg/m³. The mean of the two highest recorded liquid and vapor densities is 536. The lowest coexisting liquid mass density they report is 602 kg/m$^3$ giving $\rho_{pa}(T_c) = 0.507$. The highest vapor mass density they can observe near $T_c$ is 470 kg/m$^3$ which corresponds to $\rho_{pb} = 0.395$. The line of critical states connects these two points.

We can calculate the characteristic bond-length that defines the bonded-cluster percolation transition from the hard-sphere model, i.e. in a first-order perturbation approximation, if the structure is not perturbed significantly by the attraction. From an EXCEL power-law trendline parameterization of the extended volume percolation transition density as a function of cluster-length $\lambda$ from table I in the paper of Heyes et al. [7], an inversion gives

$$\lambda = 1 + 0.0412 \, \rho^{-1.617} \qquad (9)$$

After conversion to L-J units, substituting $\rho$ for the highest vapor density of Gilgen et al. at the critical temperature, we obtain $\lambda_{pa}(T_c) = 1.185$, which is in the anticipated range of the Lennard-Jones attraction; slightly greater than the distance of maximum attractive force which, from the first derivative of equation (2), is $(13/7)^{1/6} = 1.1086$.

To complete the construction of the phase diagram in **Figure 7**, we need the two percolation densities $\rho_{pe}$ and $\rho_{pa}$ of the soft-sphere fluid which is the high-temperature limit of the L-J fluid. The density of PE for purely repulsive soft-spheres $\rho_{pe}$ (ss) = 0.08 is obtained from Heyes et al.[7], and is essentially the same value as the hard sphere $\rho_{pe}$ .[13] The density of the soft-sphere available volume percolation density $\rho_{pa}$(ss) = 0.48 is obtained from a determination from discontinuity in rheological properties.[16] The equilibrium fluid freezing density of the high-temperature limit is obtained from Hoover et al. for the soft-sphere model[17], $\rho_f$(ss) = 1.51. These three state points are the limiting high-temperature bounds of argon's three fluid phases as $\beta = \varepsilon/k_B T \rightarrow 0$.

The bonded cluster percolation transition (line PB) clearly must become increasingly weaker with temperature and eventually non-existent, probably before it crosses the excluded volume percolation transition (line PE) at the much lower density. It is not presently known what the extent and manifestations of the excluded volume percolation transition are at low temperatures. There is a possibility of a second mesophase bounded by loci PE and PB.

## WATER AND STEAM

All liquids should exhibit both supercritical higher-order percolation transitions which, at a sufficient low temperature, will become coincident upon a coexisting vapor and liquid-state critical densities, i.e. at the line of critical states. In **Figure 8** we have plotted the p ($\rho$,T) experimental data, also tabulated with 6-figure accuracy, published for water. The results are the same as seen for model square–well molecules (**Figure 4**) and for argon (**Figure 5**). The isotherms clearly show the change in slope on either side of the liquid-vapor coexistence, and a horizontal water-water vapor critical coexistence tie-line.

Bernebei et al.[18] have recently reported the discovery of a structural phase change, claimed by the authors to be the evidence for the existence of a "percolation transition" in supercritical water. These authors refer to a supercritical "percolation line" accompanied by a "change of symmetry", dividing "gas-like" and "liquid-like" structures respectively. From only 4 data points, two of which are at very high densities (750kg/m$^3$) and off the scale they are unable to locate the percolation threshold that they refer to. The percolation transitions, as shown in **Figure 8,** are likely to be the originating thermodynamic explanation of the structural changes seen by Bernebei et al. Further experimental investigations may reveal structural changes accompanying the lower-density bonded-cluster percolation transition (PB) when a cluster of hydrogen-bonded molecules first begins to span the system.

The liquid phase that we call "water" is now seen to extend into the supercritical region all the way to a high temperature with no limit. It is not a critical isotherm that bounds the water phase as commonly assumed to distinguish water from steam. Water as a phase spans all temperatures. Below $T_c$ it is bounded by the liquid-vapor coexistence line, above $T_c$ the low-density bound is the line of the available volume percolation density $\rho_{pa}$. Note also in **Figure 8** that two distinct steam phases can now be identified on either side of the bonded-cluster percolation transition line (PB) which are designated "steam I" and steam II, the latter being a supercritical mesophase.

**CONCLUSIONS**

Boundaries of the liquid state have not previously been properly defined. One often sees arbitrary lines drawn with no scientific status, in school physics text books, and indeed also in 'popular science' journals such as the recent article by McMillan and Stanley [20]. Some authors use the critical isotherm to define the supercritical liquid boundary, whilst others the critical isobar (see Figure 1 in reference 20); these lines have no thermodynamic status as phase boundaries. Here we now see the defining boundary line is neither the critical isotherm, nor the critical isobar, but the available volume percolation transition loci PA in the density p-T surface, i.e. $\rho_{pa}(p,T)$. The liquid phase spans all temperatures from a metastable amorphous ground state at absolute zero,[1] to the highest temperature limit that can be attained with molecular integrity. In the case of a classical model Hamiltonian, such as the LJ model, the high temperature limit is infinite. The boundaries of any liquid phase can be defined thermodynamically in terms of first and second-order phase transitions on Gibbs surfaces.

Universality is a concept that frequently arises in the general description of critical properties of disordered systems.[4,5] Analytical solutions with critical exponents are found to be the same in different dimensions for mean-field model fluids[21]. Such models, however, may not be applicable to real fluids with continuous Hamiltonians that exhibit percolation transitions. In 1D, for example, percolation transitions don't exist. In 2D there is only one percolation transition for extended and available volume of hard-discs, i.e. quite different from 3D. Thus we expect the physics of gas-liquid critical condensation to be non-universal in the sense that the thermodynamic description will be different in lower dimensions as percolation phenomena are strongly dimension dependent. Thus we conclude that there is no multi-dimensional universality, as it applies, for example, to criticality in model lattice gases, in the thermodynamic description of gas-liquid critical phenomena.

The present thermodynamic description explains why no one has ever obtained a critical density by a direct experimental measurement. Experimentalists, being unable to observe a critical point density directly, but resorted to extrapolation using the rectilinear diameter law.[22] Now we see that critical densities, as reported in the in the literature, are actually a mean value of the percolation densities $\rho_{pb}$ and $\rho_{pa}$ at $T_c$. It is interesting to note, that Andrews originally determined a critical temperature by observing a temperature above which the coexistence line was no longer horizontal.[2] That was before van der Waals! Since 1873, the two basic assumptions to obtain a critical density were first, an assumption of some universal form of a reduced equation-of-state in the vicinity of $T_c$, and second, the law of rectilinear diameters which enabled the prediction of a hypothetical 'critical density' from the measurable vapor and liquid coexistence densities.

The same applies to previous simulation studies from which critical densities have been extracted by extrapolation using near-critical equation-of-state universality, and a law of rectilinear diameters, for example the square-well studies[9,10] in Figure 4. The Gibbs ensemble simulation of argon by Wilding[23] is another typical example. In figure 7 of Wilding's paper[23], one can see that using these Monte Carlo simulations the lowest obtainable liquid density in coexistence with a vapor is 0.485. This is in good agreement with the minimum coexisting liquid density that Gilgen et al could actually measure[15] (0.507 in LJ units: Figure7), and notably close to $\rho_{pa}$ for hard spheres[13] (0.537). In common with the above mentioned square-well critical density extractions; to obtain the critical parameters Wilding usurps, *a priori*, the existence of a critical point on the $\rho(p,T)$ surface with exponents described by a theoretical universality in the description of energy and density fluctuations. He then uses the law of rectilinear diameters (see equation (3.10) and figure 7 of Wilding's paper) to estimate the critical point parameters of the Lennard-Jones fluid.

It is now 25 years since percolation transitions were shown to be related to abrupt changes in linear transport coefficients and rheological properties of fluids.[16,24-26] Only recently, however, have studies of the percolation thresholds in simple hard-sphere model fluids [7,13] revealed a connection between percolation phenomena and 2nd order thermodynamic phase transitions, resulting in the present and previous[1] observation of sub phases in the Gibbs state-function surfaces of supercritical fluids. There are host of recent papers[20,27,28] on supercritical "lines", reporting the discovery of various discontinuities in dynamical properties, transport coefficients, maxima in second-order properties, or correlation lengths etc. etc. All these 'mysterious lines', with no attendant thermodynamic explanations, may be described by the second-order thermodynamic percolation transition loci in the Gibbs p,T surface that give rise to supercritical sub-phases.

Interestingly, the little graphic presented within the abstract of the paper of Brazhkin et al.[27] shows three lines of second-order property maxima stemming from their critical point in the p-T plane. These three lines can be identified with thermodynamic percolation transition loci, shown here for argon in Figure 5, also in the p-T plane. The correlation length $\zeta(T)$-max locus is near PB, the $C_p$-max locus is near PA, and a relatively flat $\alpha_p$-max locus is intermediate between PA and PB.

In conclusion, a longstanding conundrum is resolved. The hypothetical critical point of van der Waals has one phase, and, if $(dp/dV)_T = 0$, only one degree of freedom, in an apparent violation of the Gibbs phase rule. By analogy with the triple point, we have found $T_c, p_c$ is actually a "double point" in the p-T plane. A critical event is caused by an intersection of two percolation transition pressure loci. At $T_c, p_c$, each intermediate state point on the Gibbs surface corresponds to a different density, and since $(Vdp)_T = d\mu = 0$ there is a connecting line of states at $T_c, p_c$ of constant Gibbs chemical potential ($\mu$), along the line of critical states. The $2^{nd}$-order percolation transitions on intersection in the p-T plane, give rise to a $1^{st}$-order phase transition with two coexisting phases and a single degree-of-freedom, now in compliance with the Gibbs phase rule.

## AUTHOR INFORMATION

e-mail: les.woodcock@manchester.ac.uk

## ACKNOWLEDGEMENTS

This paper was prepared under the auspices of the Korean Federation of Universities of Science and Technology (KFUST) Brain-pool Scheme; we thank the Department of Physics Kong University and host Professor Sangrak Kim.

**TABLE**

**Table 1**: Ensemble average of pressures ($p^*=p\sigma^3/\varepsilon$) of the square-well fluid ($\lambda=5$) for near-critical and supercritical isotherms obtained from N-V-T Monte Carlo simulations with N=6912 and periodic boundary conditions. The range of uncertainty in pressure is within $\pm$ 5%; for T* =40 the errors may be as high as $\pm$ 10% in the two-phase region of the slightly sub-critical isotherm.

| Density | T*=40 | T*=45 | T*=50 | T*=60 | T*=70 | T*=80 |
|---------|-------|-------|-------|-------|-------|-------|
| 0.05 | 2.136 | 2.154 | 2.105 | 2.633 | 3.292 | 3.773 |
| 0.10 | 2.531 | 3.065 | 3.498 | 4.584 | 5.447 | 7.218 |
| 0.15 | 2.353 | 3.295 | 4.518 | 6.245 | 8.756 | 10.55 |
| 0.20 | 1.413 | 2.958 | 5.336 | 8.788 | 12.36 | 14.06 |
| 0.25 | 0.342 | 3.147 | 5.553 | 9.492 | 15.45 | 18.05 |
| 0.30 | -0.609 | 0.828 | 5.523 | 11.45 | 16.90 | 22.90 |
| 0.35 | -0.740 | 1.198 | 6.255 | 15.01 | 22.32 | 29.21 |
| 0.40 | -0.325 | 3.857 | 8.432 | 19.36 | 30.05 | 37.49 |
| 0.45 | 0.362 | 3.987 | 10.83 | 23.89 | 38.21 | 48.53 |
| 0.50 | 1.979 | 6.383 | 15.23 | 35.77 | 48.77 | 64.28 |
| 0.55 | 5.130 | 13.58 | 21.19 | 47.32 | 65.96 | 85.92 |
| 0.60 | 10.72 | 23.75 | 35.37 | 64.05 | 86.98 | 110.9 |
| 0.65 | 20.66 | 39.45 | 56.79 | 85.56 | 111.7 | 143.3 |
| 0.70 | 35.45 | 62.56 | 83.14 | 113.8 | 154.5 | 192.4 |

**CAPTIONS TO FIGURES (6)**

**Figure 1**         (a) p-ρ isotherms of van der Waals (VDW) equation
                         (b) p-ρ isotherms of augmented van der Waals (AVDW) equation
                         (c) p-ρ isotherms of square-well ($\lambda=5$) fluid from MD simulations

**Figure 2**   Mean pressures ($p^*=p\sigma^3/\varepsilon$) from N-V-T MC simulations of the square-well fluid ($\lambda=5$) various isotherms; red T*= 40, orange T*= 45, lime T*= 50, green T*= 60, blue T*=70, violet T*=80; the vertical dashed lines are the hard-sphere fluid percolation transition densities $\rho_{pe}$ and $\rho_{pa}$.

**Figure 3**   Pressure difference ($\Delta p= p_{hs}-p_{sw}$) between SW fluid ($\lambda=5$) and the hard sphere fluid for 4 supercritical isotherms (black lines); the estimated $T_c^*$ = 42; the red line is the pressure $p_{hs}$ ($p\sigma^3/k_BT$) of the hard-sphere fluid. The quadratic equations are the EXCEL trendlines. The vertical dashed lines are the hard-sphere percolation densities $\rho_{pe}$ and $\rho_{pa}$.

**Figure 4**   Coexistence curves of square-well fluids with limiting densities; curves for $\lambda=1.25$ to 2.0 Vega et al. [9]; open circles are the densities of Elliott and Hu [10]; densities at $\lambda=3.0$ from Benavides et al [12]; densities and curves for $\lambda=1.005$ and $\lambda=5$ are present results. The upper dashed line is the available volume percolation transition loci ($\rho_{pa}=0.537$), and the lower horizontal dashed lines are the extended volume

percolation transition loci, of the hard-sphere reference fluid; $\rho_{pb}(\lambda)$ is the high-T limit, $\lambda$-dependent, bonded-cluster percolation density at $T_c$ from equation 6 in the text.

**Figure 5** Experimental pressure-density isotherms for liquid argon; the data is plotted directly from the tabulations in Table 1 of Gilgen et al.[14] for the critical isotherm (150.7K) and 7 supercritical isotherms. The straight lines have been superimposed to highlight the percolation transitions determined by the discontinuity in $(dp/d\rho)_T$.

**Figure 6** Projection in the p-T plane of the experimental liquid-vapor coexistence surface for argon[15] together with the loci of the supercritical percolation transitions; obtained from the thermodynamic data tables of Gilgen et al.[14] for the 7 supercritical isotherms shown in Figure 5.

**Figure 7** Phase diagram of argon, excluding crystalline phases, reduced to Lennard-Jones units; the red dots show the coexisting densities from the experimental measurements of Gilgen et al.[14,15]; percolation lines are from figure 6; the densities of the soft-sphere percolation transitions PE and PA in the high-temperature limit ($\beta$=0), and the soft-sphere freezing density (black dots) are obtained from references (7), (16) and (17) respectively.

**Figure 8** Critical and supercritical isotherms of water and steam plotted directly from the numerical p-V-T data tabulations given Wagner and Kretzschnar [18]; the critical point of water is 376°C. The straight lines have been superimposed to show the loci of the percolation transitions at the change in slope of the isotherms. The two black stars on the 400°C isotherm are the state points investigated experimentally by Bernabei et al.[19]

**FIGURES 1- 6**
Figure 1

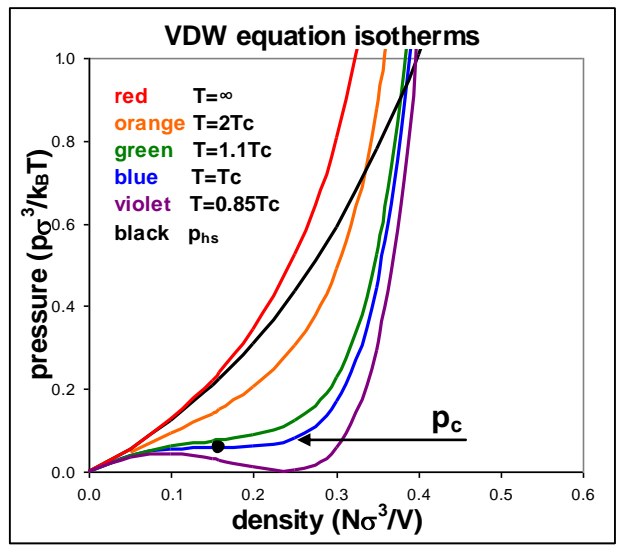

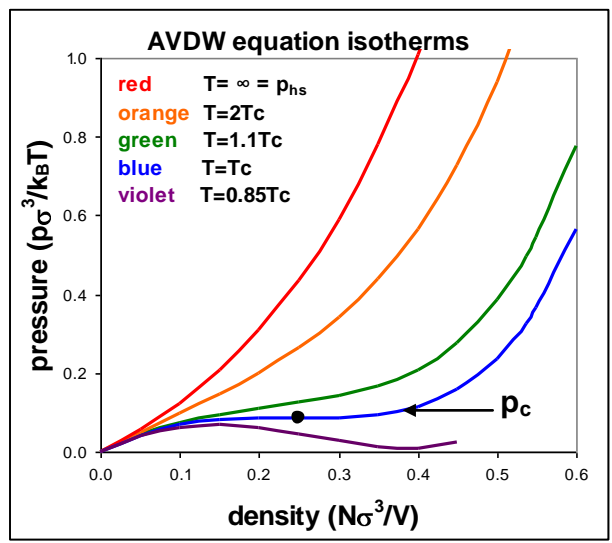

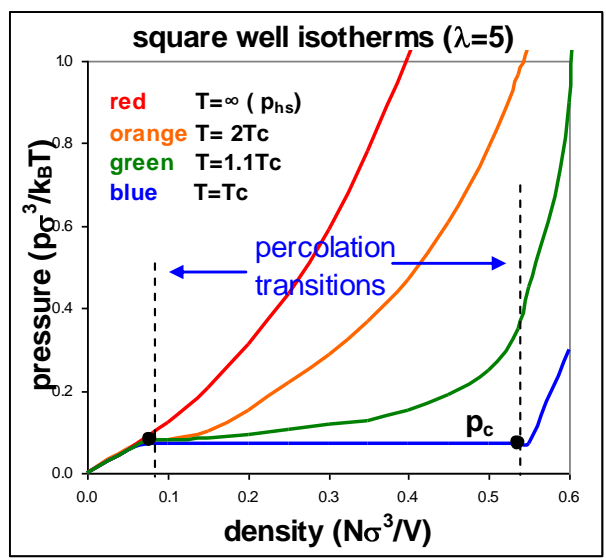

Figure 2

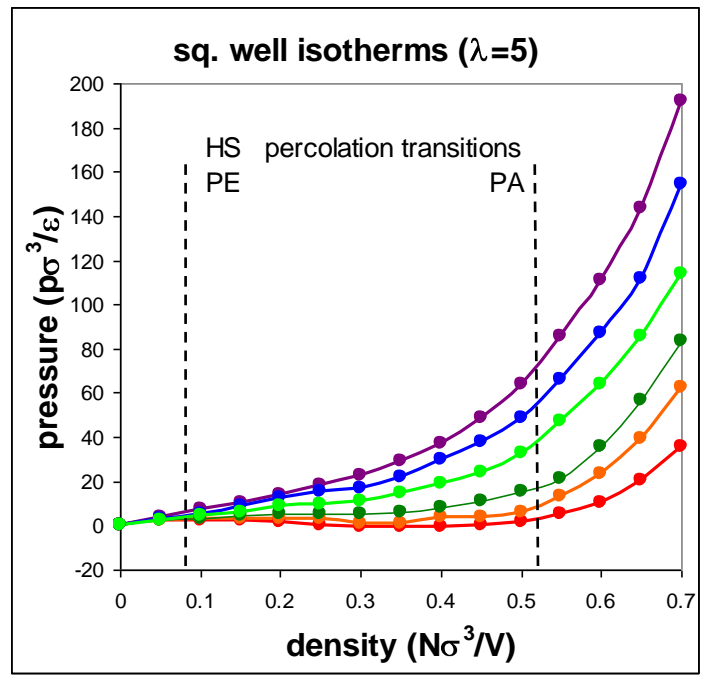

Figure 3

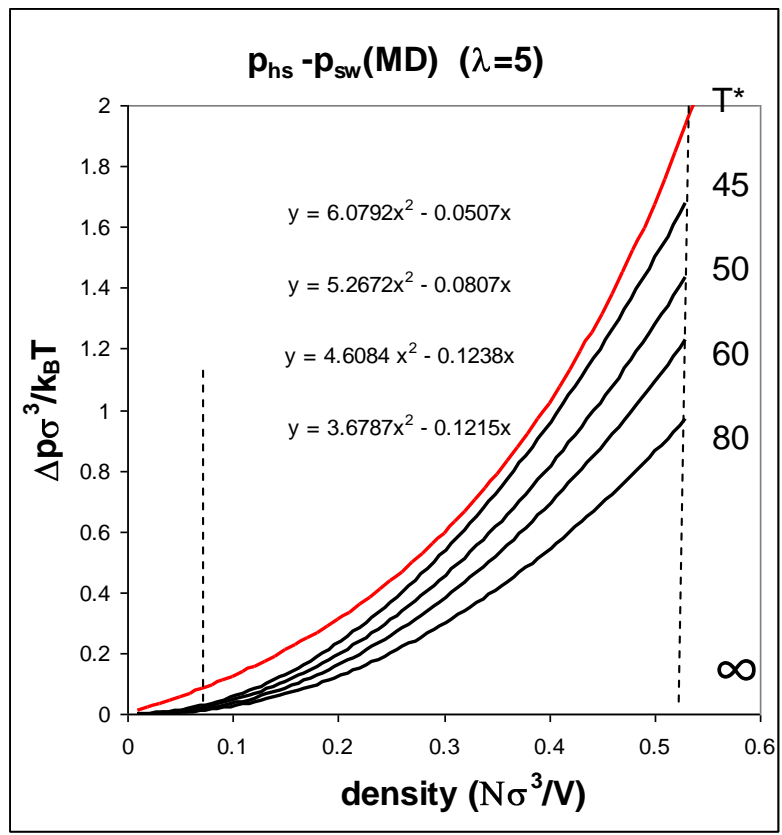

Figure 4

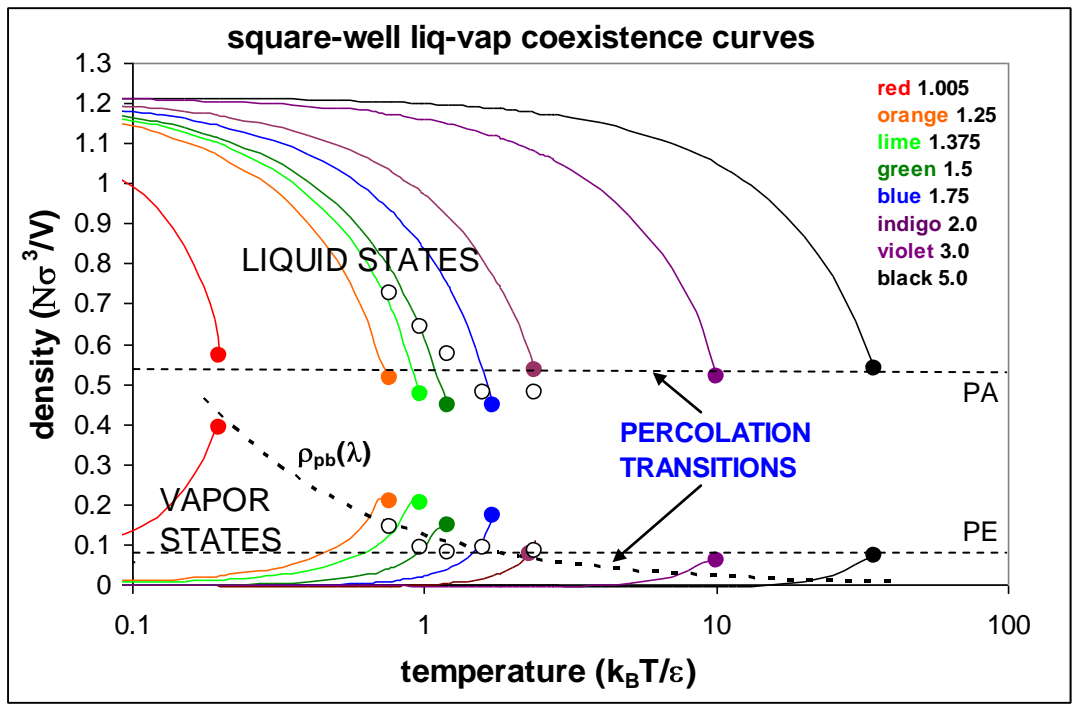

Figure 5

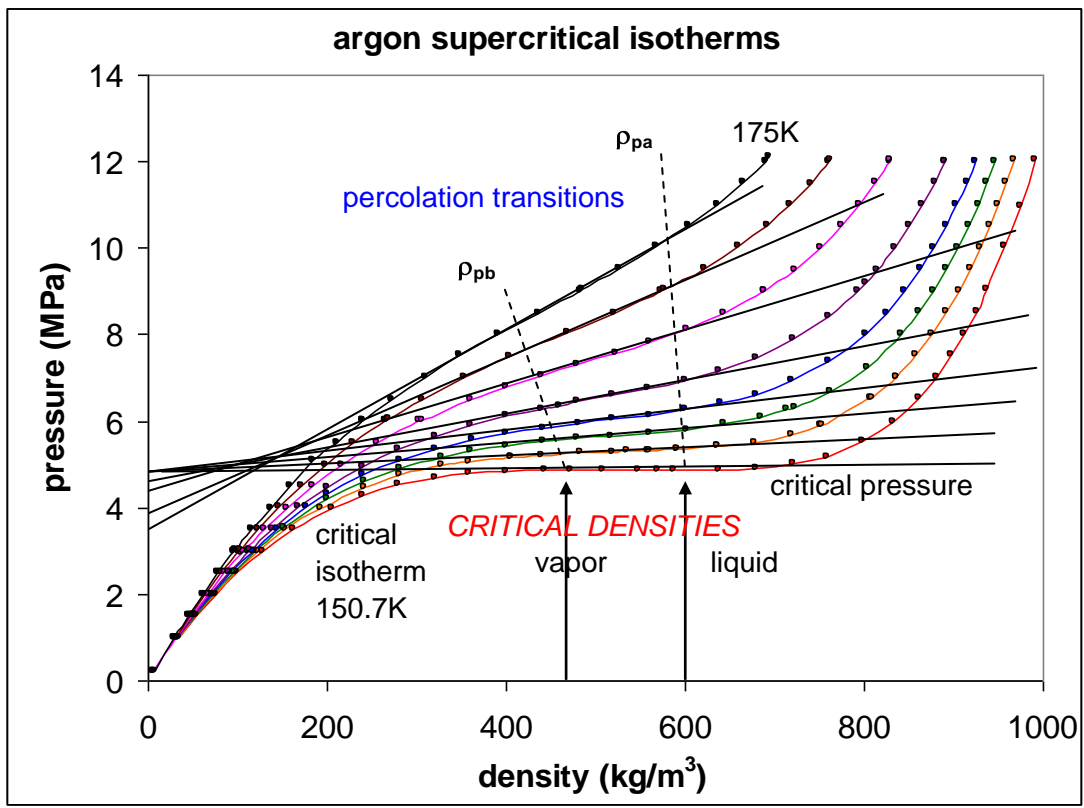

Figure 6

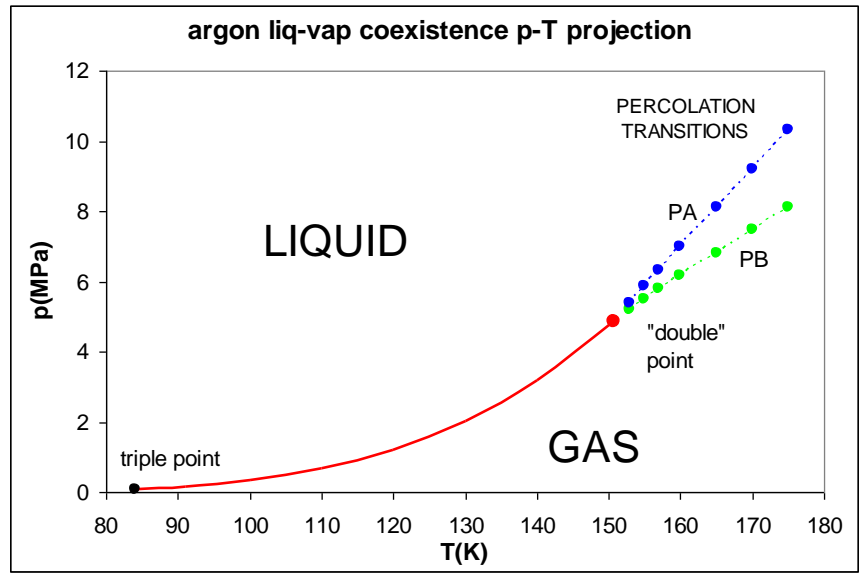

Figure 7

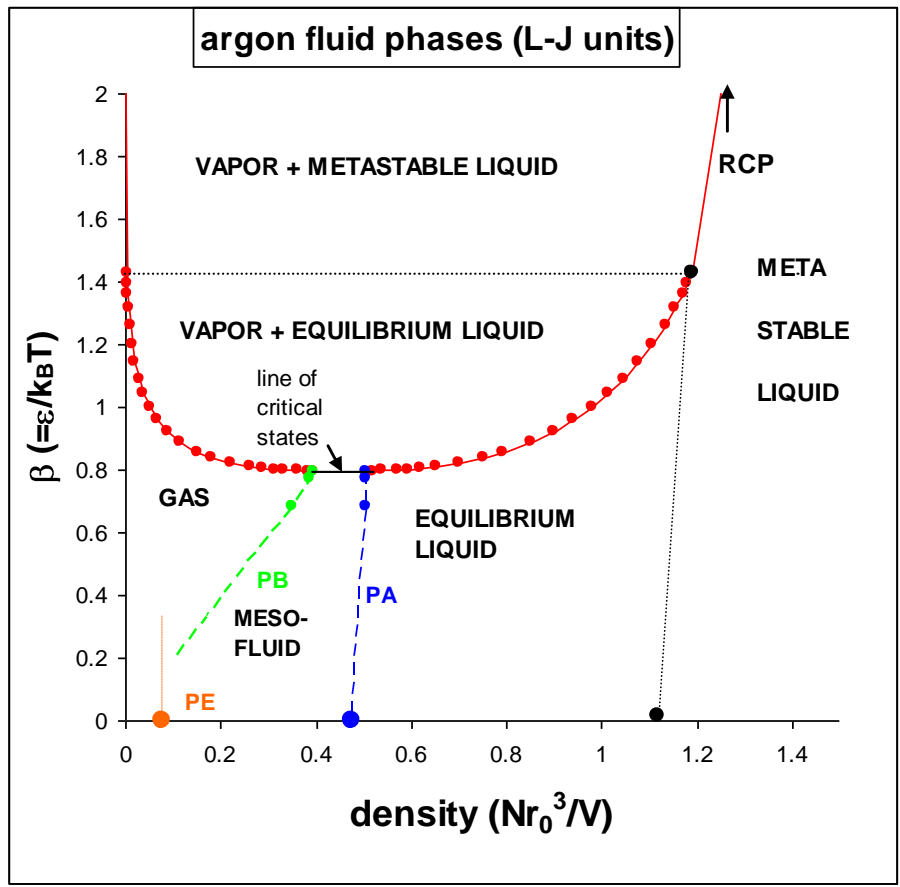

Figure 8

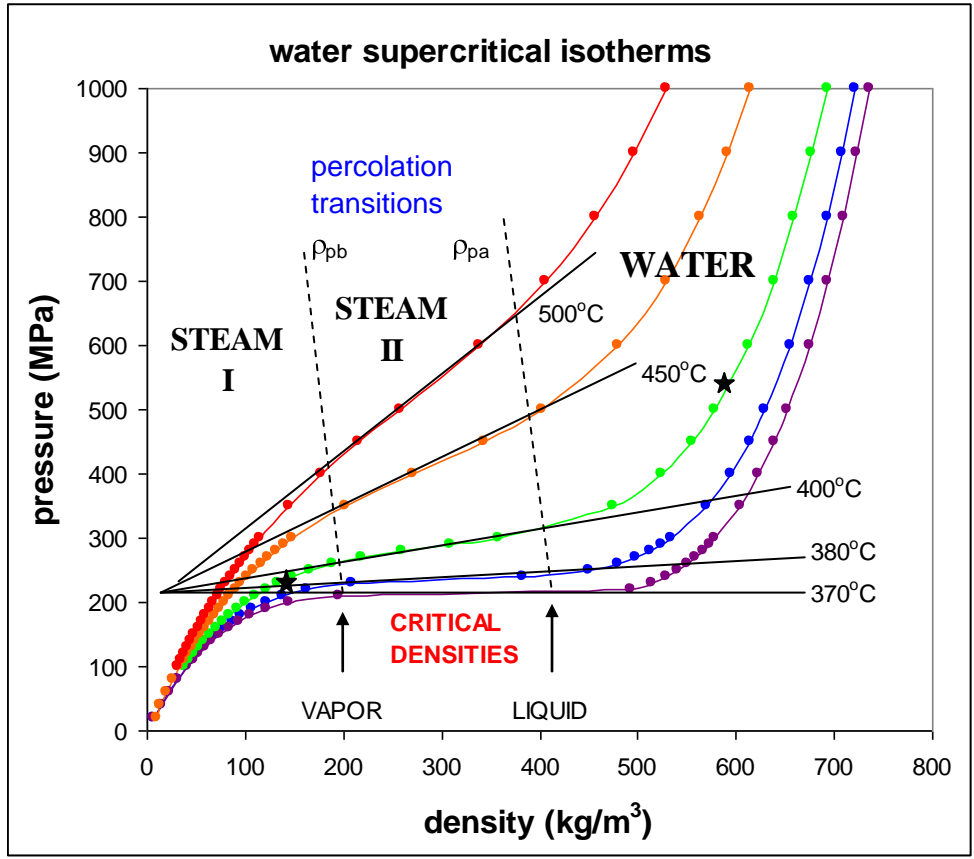

**ABSTRACT GRAPHIC**

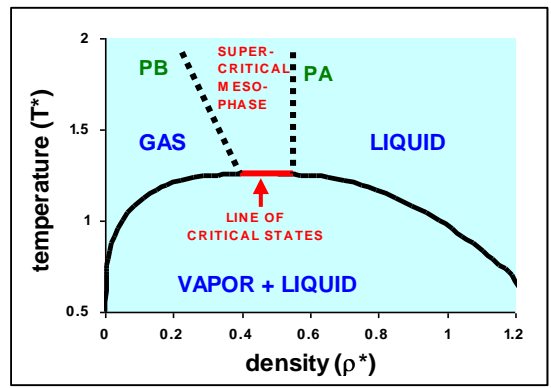